\documentclass[aps,prl,twocolumn,groupedaddress,showpacs]{revtex4}
\usepackage{graphicx,amsmath,amsfonts,amssymb,mathrsfs,bbm}

\begin{document}

\title{Mapping Out $SU(5)$ GUTs with Non-Abelian Discrete Flavor Symmetries}

\author{Florian Plentinger}
\email[]{fplentinger@physik.uni-wuerzburg.de}
\author{Gerhart Seidl}
\email[]{seidl@physik.uni-wuerzburg.de}
\affiliation{Institut f\"ur Physik und Astrophysik, Universit\"at
W\"urzburg, Am Hubland, D 97074 W\"urzburg, Germany}

\date{\today}

\begin{abstract}
We construct a class of supersymmetric $SU(5)$ GUT models that
produce nearly tribimaximal lepton mixing, the observed quark mixing
matrix, and the quark and lepton masses, from discrete non-Abelian
flavor symmetries. The $SU(5)$ GUTs are formulated on five-dimensional throats in the flat limit and the neutrino masses become
small due to the type-I seesaw mechanism. The discrete non-Abelian flavor
symmetries are given by semi-direct products of cyclic groups that are
broken at the infrared branes at the tip of the throats. As a result, we obtain $SU(5)$ GUTs that provide a combined description of
non-Abelian flavor symmetries and quark-lepton complementarity.
\end{abstract}

\maketitle

One possibility to explore the physics of grand unified theories (GUTs)
\cite{SU5,Pati:1974yy} at low energies is to analyze the neutrino
sector. This is due to the explanation of small neutrino masses
via the seesaw mechanism
\cite{typeIseesaw,typeIIseesaw}, which is naturally incorporated in
GUTs. In fact, from the perspective of quark-lepton unification, it is
interesting to study in GUTs the drastic differences between
the masses and mixings of quarks and leptons as revealed by current
neutrino oscillation data.

In recent years, there have been many attempts to reproduce a
tribimaximal mixing form \cite{Harrison:1999cf} for the leptonic
Pontecorvo-Maki-Nakagawa-Sakata (PMNS) \cite{PMNS} mixing matrix
$U_{\text{PMNS}}$ using non-Abelian discrete
flavor symmetries such as the tetrahedral \cite{A4} and double
(or binary) tetrahedral \cite{T'} group
\begin{equation}
A_4\simeq Z_3\ltimes(Z_2\times Z_2)\quad\text{and}\quad T'\simeq Z_2\ltimes Q,
\end{equation}
where $Q$ is the quaternion group of order eight, or \cite{delta27}
\begin{equation}
\Delta(27)\simeq Z_3\ltimes (Z_3\times Z_3),
\end{equation}
which is a subgroup of $SU(3)$ (for reviews see, e.g., Ref.~\cite{Ma:2007ia}). Existing models, however, have generally difficulties to predict also the observed fermion
mass hierarchies as well as the Cabibbo-Kobayashi-Maskawa (CKM) quark
mixing matrix $V_{\text{CKM}}$ \cite{CKM}, which applies especially to GUTs (for very recent examples, see Ref.~\cite{discreteGUTs}). Another approach, on the other hand, is
offered by the idea of quark-lepton complementarity (QLC), where
the solar neutrino angle is a combination of maximal mixing
and the Cabibbo angle $\theta_\text{C}$ \cite{qlc}. Subsequently, this
has, in an interpretation of QLC
\cite{Plentinger:2006nb,Plentinger:2007px}, led to a machine-aided survey of several thousand lepton flavor models
for nearly tribimaximal lepton mixing \cite{Plentinger:2008up}.

Here, we investigate the embedding of the models found in Ref.~\cite{Plentinger:2008up} into five-dimensional (5D)
supersymmetric (SUSY) $SU(5)$ GUTs. The hierarchical pattern of quark
and lepton masses, $V_\text{CKM}$, and nearly
tribimaximal lepton mixing, arise from the local breaking of
non-Abelian discrete flavor symmetries in the extra-dimensional
geometry. This has the advantage that the scalar sector
of these models is extremely simple without the need for a
vacuum alignment mechanism, while offering an intuitive geometrical interpretation
of the non-Abelian flavor symmetries. As a consequence,
we obtain, for the first time, a realization of non-Abelian
flavor symmetries and QLC in $SU(5)$ GUTs.

\begin{figure}
 \includegraphics*[bb = 155 645 460 710,width=8.5cm]{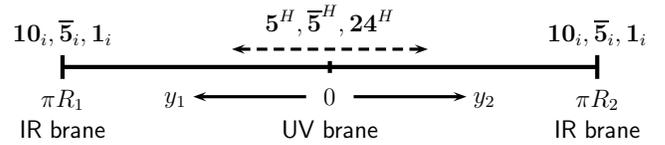}
 \caption{SUSY $SU(5)$ GUT on two 5D intervals or throats. The zero
 modes of the matter fields ${\bf 10}_i,\overline{\bf 5}_i,$ and ${\bf 1}_i$,
 are symmetrically located at $y_1=\pi R_1$ and $y_2=\pi R_2$, whereas
 the Higgs hypermultiplets ${\bf 5}^H,\overline{\bf 5}^H,{\bf
 24}^H$, and the gauge supermultiplet, propagate
 freely in the two throats.
\label{fig:throats}}
 \end{figure}

We will describe our models by considering a specific minimal
realization as an example. The main features of this
example model, however, should be viewed as generic and representative
for a large class of possible realizations. Our model is given by a SUSY $SU(5)$
GUT in 5D flat space, which is defined on two 5D intervals that have been
glued together at a common endpoint. The geometry and the location of the 5D
hypermultiplets in the model is depicted in FIG.~\ref{fig:throats}.
The two intervals constitute a simple example for
a two-throat setup in the flat limit (see, e.g.,
Refs.~\cite{Cacciapaglia:2006tg,Agashe:2007jb}), where the two 5D
intervals, or throats, have the lengths $\pi R_1$
and $\pi R_2$, and the coordinates $y_1\in[0,\pi R_1]$ and
$y_2\in[0,\pi R_2]$. The point at $y_1=y_2=0$ is called ultraviolet
(UV) brane, whereas the two
endpoints at $y_1=\pi R_1$ and $y_2=\pi R_2$ will be referred to as infrared
(IR) branes. The throats are supposed to be GUT-scale sized,
i.e.~$1/R_{1,2}\gtrsim M_\text{GUT}\simeq 10^{16}\,\text{GeV}$, and
the $SU(5)$ gauge supermultiplet and the
Higgs hypermultiplets ${\bf 5}^H$ and $\overline{\bf 5}^H$ propagate
as bulk fields freely on the two intervals. In usual 5D GUT models,
$SU(5)$ is broken to the standard model (SM) gauge group
$G_{\text{SM}}=SU(3)_c\times SU(2)_L\times U(1)_Y$ by boundary
conditions \cite{Kawamura:2000ev}. In contrast to this, we suppose
that $SU(5)$ is spontaneously broken to $G_{\text{SM}}$ by a
${\bf 24}^H$ bulk Higgs hypermultiplet propagating in the two
throats that acquires a vacuum expectation value pointing in the hypercharge
direction $\langle {\bf
  24}^H\rangle\propto\text{diag}(-\frac{1}{2},-\frac{1}{2},\frac{1}{3},\frac{1}{3},\frac{1}{3})$. Therefore, we have the usual $SU(5)$ explanation of SM quantum
numbers and charge quantization.

The zero modes of all three generations of fermion superfields are
assumed to be symmetrically localized at both IR branes of the
throats. This symmetric trapping of zero modes at the tip of the
throats could be achieved as in Ref.~\cite{Cacciapaglia:2006tg} (see
also Ref.~\cite{Kaplan:2001ga}) by introducing suitable bulk fermion
masses in the throats. As a result of this localization, we describe the
fermion zero modes in the language of 4D
$N=1$ SUSY. In doing so, it is assumed that some 4D $N=2$
SUSY (which is equivalent to minimal 5D SUSY) is locally broken down
to 4D $N=1$ SUSY at the UV/IR branes.

The quark and lepton zero modes are contained in the $SU(5)$ matter chiral superfields ${\bf 10}_i$ and $\overline{\bf 5}_i$, where $i=1,2,3$ is the generation
index. To obtain small neutrino masses via the type-I seesaw
mechanism \cite{typeIseesaw}, we introduce three right-handed $SU(5)$ singlet neutrino superfields
${\bf 1}_i$. The 5D Lagrangian for the Yukawa
couplings of the zero mode fermions then reads
\begin{eqnarray}\label{eq:5DLagrangian}
\mathcal{L}_{5D}&=&\int
d^2\theta{\big [}\delta(y_1-\pi R_1){\big(}\tilde{Y}^{u}_{ij,R_1}{\bf 10}_i{\bf 10}_j{\bf
5}^H\nonumber\\
&+&\tilde{Y}^{d}_{ij,R_1}{\bf 10}_i\overline{\bf 5}_j\overline{\bf
5}^H+\tilde{Y}^\nu_{ij,R_1}\overline{\bf 5}_i{\bf 1}_j{\bf 5}^H
+M_R\tilde{Y}^R_{ij,R_1}{\bf 1}_i{\bf 1}_j{\big)}\nonumber\\
&+&\delta(y_2-\pi R_2){\big(}\tilde{Y}^{u}_{ij,R_2}{\bf 10}_i{\bf 10}_j{\bf 5}^H+\tilde{Y}^{d}_{ij,R_2}{\bf 10}_i\overline{\bf 5}_j\overline{\bf
5}^H\nonumber\\
&+&\tilde{Y}^\nu_{ij,R_2}\overline{\bf 5}_i{\bf 1}_j{\bf 5}^H
+M_R\tilde{Y}^R_{ij,R_2}{\bf 1}_i{\bf 1}_j{\big )}+\text{h.c.}{\big]},
\end{eqnarray}
where $\tilde{Y}^x_{ij,R_1}$ and $\tilde{Y}^x_{ij,R_2}$ ($x=u,d,\nu,R$) are Yukawa
coupling matrices (with mass dimension $-1/2$) and $M_R\simeq
10^{14}\,\text{GeV}$ is the $B-L$ breaking scale. In the four-dimensional (4D) low energy effective theory, $\mathcal{L}_{5D}$ gives rise to the 4D Yukawa couplings
\begin{eqnarray}\label{eq:4DLagrangian}
\mathcal{L}_{4D}&=&\int
d^2\theta{\big [}Y^{u}_{ij}{\bf 10}_i{\bf 10}_j{\bf
5}^H+Y^{d}_{ij}{\bf 10}_i\overline{\bf 5}_j\overline{\bf 5}^H\nonumber\\
&+&Y^\nu_{ij}\overline{\bf 5}_i{\bf 1}_j{\bf 5}^H
+M_RY^R_{ij}{\bf 1}_i{\bf 1}_j+\text{h.c.}{\big]},
\end{eqnarray}
where $Y^x_{ij}=(M_* \pi
R)^{-1/2}(\tilde{Y}^x_{ij,R_1}+\tilde{Y}^x_{ij,R_2})$ are the
dimensionless Yukawa coupling matrices of the low-energy theory,
$M_*\simeq(M_{\text{Pl}}^2R_{1,2}^{-1})^{1/3}$ is the fundamental scale, and
$M_{\text{Pl}}\simeq 10^{19}\,\text{GeV}$ the usual 4D Planck
scale. Note from Eq.~(\ref{eq:4DLagrangian})
that small neutrino masses are generated via the canonical type-I seesaw
mechanism after integrating out the ${\bf 1}$s. A crucial property of the 4D Yukawa couplings $Y^x_{ij}$, which
we will exploit later, is that they receive contributions from both IR branes of the throats.

Now, we extend the gauge group to $SU(5)\times G_F$,
where $G_F$ is a discrete non-Abelian flavor symmetry group. We assume
that $G_F$ is a semi-direct product of two flavor groups $G_A$ and $G_B$, i.e.~$G_F=G_A\ltimes
G_B$. Here, $G_A$ is taken to be a direct product of $Z_n$ symmetries,
i.e. $G_A=Z_{n_1}\times Z_{n_2}\times\dots\times Z_{n_m}$, where $m$ is the number of $Z_n$ factors and the $n_k$
($k=1,2,\dots,m$) may be different. Under $G_A$, we assign to each
generation $i$ the charges
\begin{eqnarray}\label{eq:charges}
{\bf 10}_i&\sim&(p^i_1,p^i_2,\dots,p^i_m),\nonumber\\
\overline{\bf 5}_i&\sim&(q_1^i,q^i_2,\dots,q^i_m),\\
{\bf 1}_i&\sim&(r^i_1,r^i_2,\dots,r^i_m)\nonumber,
\end{eqnarray}
where the $j$th entry in each row vector denotes the
$Z_{n_j}$ charge of the representation. In the 5D theory, we suppose that the group $G_A$ is
spontaneously broken by singly charged flavon fields located at the IR
branes. The Yukawa coupling matrices of quarks and leptons are then generated by the Froggatt-Nielsen mechanism \cite{Froggatt:1978nt}.

\begin{figure}
 \includegraphics*[width=6.7cm]{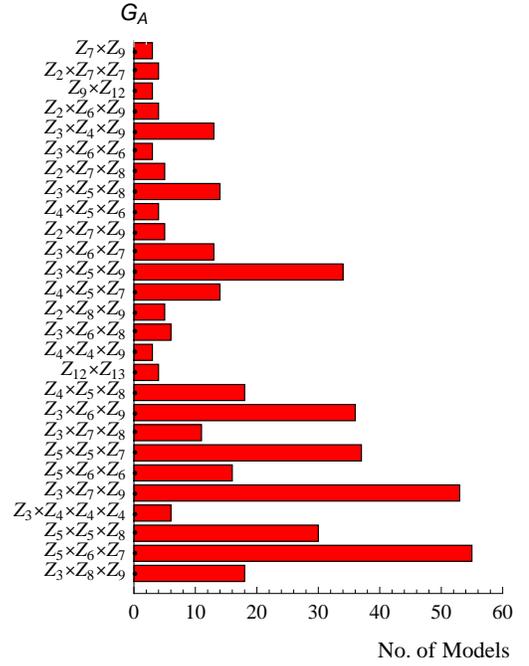}
 \caption{Overview of $SU(5)\times G_A$ models for varying flavor
 group $G_A$. Each model yields an excellent fit to quark and lepton masses, $V_\text{CKM}$, and
 nearly tribimaximal lepton mixing. All models produce a reactor
 neutrino angle $\ll 1^\circ$ and a normal neutrino
 mass hierarchy. The graph summarizes about
 $4\times 10^2$ realistic GUTs.
\label{fig:models}}
 \end{figure}

Applying a straightforward generalization of the flavor group space scan in
Ref.~\cite{Plentinger:2008up} to the $SU(5)\times G_A$ representations in
Eq.~(\ref{eq:charges}), we find a large number of about $4\times 10^2$ flavor
models that produce the hierarchies of quark and lepton masses and yield
the CKM and PMNS mixing angles in perfect agreement with current data. A distribution of these models as a
function of the group $G_A$ for increasing group order is shown in FIG.~\ref{fig:models}.
The selection criteria for the flavor models are as follows: First, all
models have to be consistent with the quark and charged lepton mass ratios
\begin{eqnarray}
m_u:m_c:m_t&=&\epsilon^6:\epsilon^4:1,\nonumber\\
m_d:m_s:m_b&=&\epsilon^4:\epsilon^2:1,\label{eq:quark+leptonmasses}\\
m_e:m_\mu:m_\tau&=&\epsilon^4:\epsilon^2:1,\nonumber
\end{eqnarray}
and a normal hierarchical neutrino mass spectrum
\begin{equation}
 m_1:m_2:m_3=\epsilon^2:\epsilon:1,\label{eq:neutrinomasses}
\end{equation}
where $\epsilon\simeq\theta_\text{C}\simeq 0.2$ is of the order of the Cabibbo angle. Second, each model has to reproduce the CKM angles
\begin{equation}\label{eq:CKM}
V_{us}\sim\epsilon,\quad V_{cb}\sim\epsilon^2,\quad V_{ub}\sim\epsilon^3,
\end{equation}
as well as nearly tribimaximal lepton mixing at 3$\sigma$ CL with an extremely
small reactor angle $\lesssim 1^\circ$. In performing the group space
scan, we have restricted ourselves to groups $G_A$ with orders roughly
up to $\lesssim 10^2$ and FIG.~\ref{fig:models} shows only groups
admitting more than three valid models. In FIG.~\ref{fig:models}, we
can observe the general trend that with
increasing group order the number of valid models per group generally
increases too. This rough observation, however, is modified by a large
``periodic'' fluctuation of the number of models, which possibly singles out
certain groups $G_A$ as particularly interesting. The highly populated
groups would deserve further systematic investigation, which is, however, beyond the scope of this paper.

From this large set of models, let us choose the group $G_A=Z_3\times
Z_8\times Z_9$ and, in the notation of Eq.~(\ref{eq:charges}), the charge assignment
\begin{eqnarray}
&{\bf 10}_1\sim(1,1,6),\;\;{\bf 10}_2\sim(0,3,1),\;\;{\bf
10}_3\sim(0,0,0),&\nonumber\\
&\overline{\bf 5}_1\sim(1,4,2),\;\;\overline{\bf
5}_2\sim(0,7,0),\;\;\overline{\bf 5}_3\sim(0,0,1),&\\
&{\bf 1}_1\sim(2,0,6),\;\;{\bf 1}_2\sim(2,6,0),\;\;{\bf 1}_3\sim(2,0,6),&
\nonumber
\end{eqnarray}
as a showcase. The Froggatt-Nielsen mechanism then generates the Yukawa coupling textures (neglecting $\mathcal{O}(1)$ coefficients)
\begin{equation}\label{eq:quarktextures}
Y^u_{ij}\sim\left(
\begin{array}{ccc}
\epsilon^6 & \epsilon^7 & \epsilon^5\\
\epsilon^7 & \epsilon^4 & \epsilon^4\\
\epsilon^5 & \epsilon^4 & 1
\end{array}
\right),\;
Y^d_{ij}\sim\epsilon\left(
\begin{array}{ccc}
\epsilon^4 & \epsilon^3 & \epsilon^3\\
\epsilon^4 & \epsilon^2 & \epsilon^4\\
\epsilon^6 & 1 & 1
\end{array}
\right),
\end{equation}
\begin{equation}\label{eq:neutrinotextures}
Y^\nu_{ij}\sim\epsilon^3\left(
\begin{array}{ccc}
\epsilon^2 & \epsilon & \epsilon^2\\
\epsilon^2 & \epsilon &\epsilon^2\\
1&\epsilon&1
\end{array}
\right),\;
Y^R_{ij}\sim\epsilon^4\left(
\begin{array}{ccc}
1& \epsilon^2 & 1\\
\epsilon^2 & \epsilon & \epsilon^2\\
1 & \epsilon^2 & 1
\end{array}
\right),
\end{equation}
from higher-dimension operators. Observe that $G_A$ produces overall suppression factors in front of the
down quark and neutrino Yukawa coupling matrices. Only the top Yukawa
coupling is not suppressed by the flavor symmetry. Note that as long as $M_* R_{1,2}\lesssim
16\pi^2$, the top Yukawa coupling can be large without requiring strong coupling below $M_\ast$ \cite{Nomura:2001tn}. In our example, we have a
moderate $\text{tan}\,\beta\sim 10$, and since $SU(5)$ is preserved on
both throats, the charged lepton Yukawa coupling matrix $Y^e_{ij}$
satisfies $Y^e_{ij}=Y^d_{ji}$. Thus, the model exhibits $b-\tau$ unification and
the usual $SU(5)$ mass relations for the first two
generations. From Eqs.~(\ref{eq:quarktextures}) and
(\ref{eq:neutrinotextures}), we see that the model
predicts the quark and lepton mass ratios in
Eqs.~(\ref{eq:quark+leptonmasses}) and
(\ref{eq:neutrinomasses}). Realistic relations between the first two
generations of charged fermion masses may then be achieved by the
Georgi-Jarlskog mechanism \cite{Georgi:1979df}. For the quarks, the model
predicts the CKM angles in Eq.~(\ref{eq:CKM}). The leptonic sector corresponds
(up to rotations of the ${\bf 10}$s) to No.~64 in the
list of 1981 matrix sets in Ref.~\cite{Plentinger:2007px}. Hence, the
$\mathcal{O}(1)$ Yukawa coupling coefficients can be fitted such that the
PMNS mixing angles are in perfect agreement with current neutrino data at
$1\sigma$ CL. As a result, in $U_{\text{PMNS}}$, the solar, atmospheric,
and reactor angle, take the values $\theta_{12}\approx 34^\circ,\theta_{23}\approx 52^\circ,$ and
$\theta_{13}\approx 0.2^\circ$, respectively. The PMNS mixing angles describe
therefore nearly tribimaximal lepton mixing with an extremely small
reactor angle $\theta_{13}\ll 1^\circ$. At the same time, the heavy right-handed neutrino masses $m_i^R$ exhibit the hierarchical
mass ratios $m_1^R:m_2^R:m_3^R=\epsilon^2:\epsilon:1$. We suppose that
the flavor symmetries are broken at high energies such as the GUT
scale. The Cabibbo angle $\theta_\text{C}$, however, is practically
stable under renormalization group running and $V_{cb}\sim\epsilon^2$
changes by a factor less than 2 when running from the Planck
scale down to low energies \cite{Arason:1991hu}. In the neutrino sector, since the light
neutrinos have a normal hierarchical mass spectrum, renormalization
group effects have hardly any impact on the mass ratios in
Eq.~(\ref{eq:neutrinomasses}) and alter the PMNS mixing angles only by
$\ll 1^\circ$ (for a more detailed discussion and references see
Ref.~\cite{Plentinger:2007px}). Within our precision, we will therefore
neglect the modification of our predictions by renormalization group effects.

While $G_A\subset G_F$ controls the order of magnitude
of the Yukawa couplings, exact relations among the Yukawa coupling matrix elements are established by
$G_B\subset G_F$. We suppose that $G_B=G_{B_1}\times G_{B_2}\times
G_{B_3}$ is a direct product of discrete groups which act (up to
conjugation by $G_A$) on the multiplets as
\begin{equation*}
 G_{B_1}:\overline{\bf 5}_2\leftrightarrow\overline{\bf 5}_3,\quad
 G_{B_2}:{\bf 1}_1\leftrightarrow {\bf 1}_3,\quad
G_{B_3}:{\bf 10}_3\rightarrow -{\bf 10}_3.
\end{equation*}
Since the permutation symmetry $G_{B_1}$ does not commute with $G_A$
(whereas $G_{B_2}$ and $G_{B_3}$ commute with $G_A$), the
total discrete flavor symmetry group $G_F$ is
non-Abelian and given by a semi-direct
product $G_F=G_A\ltimes G_B$. The symmetries $G_{B_1}$ and $G_{B_2}$
establish the exact relations
\begin{equation}\label{eq:relations}
 Y^d_{32}=Y^d_{33},\quad
Y^\nu_{21}=Y^\nu_{23},\quad
Y^\nu_{31}=Y^\nu_{33},\quad Y^R_{11}=Y^R_{33}
\end{equation}
for the Yukawa couplings in  Eq.~(\ref{eq:4DLagrangian}). To avoid
wrong predictions for other Yukawa couplings, $G_B$
has to be broken. Recall from Eq.~(\ref{eq:5DLagrangian}) that the 4D Yukawa couplings receive
contributions from both IR branes of the throats. We therefore assume
that $G_B$ is locally broken at the IR branes as follows:
\begin{eqnarray}
 \text{at $y_1=\pi R_1$}&:& G_B\rightarrow G_{B_1},\nonumber\\
 \text{at $y_2=\pi R_2$}&:& G_B\rightarrow G_{B_2}\times G_{B_3}.
\end{eqnarray}
Taking the total contribution to the Yukawa coupling matrices that are generated by
$G_F$ at both IR branes into account, we reproduce the textures in
Eqs.~(\ref{eq:quarktextures}) and (\ref{eq:neutrinotextures}). Now,
however, the textures obey the additional exact relations in
Eq.~(\ref{eq:relations}).

\begin{figure}
\includegraphics*[width=8.8cm]{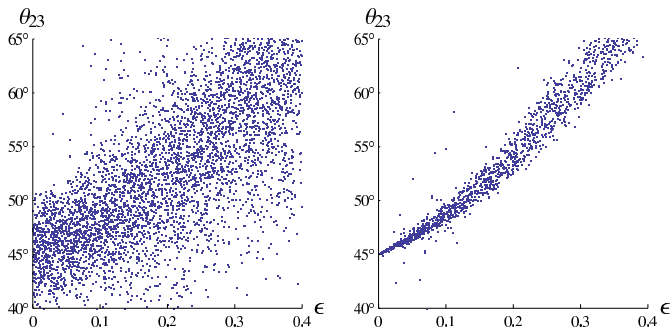}
\caption{Effect of the non-Abelian flavor symmetry on
  $\theta_{23}$ for a 10\% variation of all Yukawa couplings.
  Shown is $\theta_{23}$ as a function of $\epsilon$ for the flavor
  group $G_A$ (left) and $G_A\ltimes G_B$ (right). The right plot illustrates the exact prediction of the zeroth order term $\pi/4$ in the expansion
  $\theta_{23}=\pi/4+\epsilon/\sqrt{2}$ due to the non-Abelian nature
  of the flavor symmetry.
\label{fig:theta23}}
\end{figure}

A Monte Carlo scan of the $\mathcal{O}(1)$ Yukawa coupling
coefficients (cf.~Ref.~\cite{Plentinger:2008up}) then shows that the model satisfies the sum
rule $\theta_{23}=\pi/4+\epsilon/\sqrt{2}$ and the relation
$\theta_{13}\simeq\epsilon^2$. The
important point is that in the expression for $\theta_{23}$, the leading order term
$\pi/4$ is exactly predicted by the non-Abelian flavor symmetry
$G_F=G_A\ltimes G_B$ (see FIG.~\ref{fig:theta23}), while $\theta_{13}\simeq
\theta_{\text{C}}^2$ is extremely small due to a suppression by the square of
the Cabibbo angle.
We thus predict a deviation $\sim\epsilon/\sqrt{2}$ from
maximal atmospheric mixing, which can be tested in future neutrino
oscillation experiments such as NO$\nu$A, T2K, or a neutrino factory
\cite{Ayres:2004js}. The scan also shows that the model can, at the
same time, accommodate the
sum rule $\theta_{12}\approx\pi/4-\epsilon/\sqrt{2}$, which is the
well-known QLC relation for the solar angle. There have been attempts in the literature to
reproduce QLC in quark-lepton unified models \cite{qlcPatiSalam},
however, the model presented here is the first realization of QLC in an $SU(5)$
GUT. Although our analysis has been carried out for the CP conserving
case, a simple numerical study shows that CP violating phases (cf.~Ref.~\cite{Winter:2007yi})
relevant for neutrinoless double beta decay and leptogenesis can be
easily included as well.

Concerning proton decay, note that since $SU(5)$ is broken by
a bulk Higgs field, the broken gauge boson masses are $\simeq
M_\text{GUT}$. Therefore, all fermion zero modes can be localized
at the IR branes of the throats without introducing rapid proton
decay through $d=6$ operators. To achieve doublet-triplet splitting
and suppress $d=5$ proton decay, we may then, e.g., resort to suitable
extensions of the Higgs sector \cite{protondecay}. Moreover, although the flavor symmetry $G_F$
 is global, quantum gravity effects might require $G_F$ to be gauged
 \cite{Krauss:1988zc}. Anomalies can then be canceled by Chern-Simons
 terms in the 5D bulk.

We emphasize that the above discussion is focussed on a specific
 minimal example realization of the model. Many $SU(5)$ GUTs with
non-Abelian flavor symmetries, however, can be constructed along the
 same lines by varying the flavor charge assignment, choosing
 different groups $G_F$, or by modifying the throat geometry. A detailed analysis of these models and variations thereof will be
 presented in a future publication \cite{inpreparation}. 

To summarize, we have discussed the construction of 5D
SUSY $SU(5)$ GUTs that yield nearly tribimaximal lepton
mixing, as well as the observed CKM mixing matrix, together with the hierarchy
of quark and lepton masses. Small neutrino masses are generated
only by the type-I seesaw mechanism. The fermion masses and mixings arise from
the local breaking of non-Abelian flavor symmetries at the IR branes
of a flat multi-throat geometry. For an example realization, we have shown
that the non-Abelian flavor symmetries can exactly predict the leading
order term $\pi/4$ in the sum rule for the atmospheric mixing
angle, while strongly suppressing the reactor angle. This makes
this class of models testable in future neutrino oscillation experiments. In
addition, we arrive, for the first time, at a combined description of
QLC and non-Abelian flavor symmetries in $SU(5)$ GUTs. One main advantage of our setup with throats is that the necessary symmetry breaking can be
realized with a very simple Higgs sector and that it can be applied to
and generalized for a large class of unified models.
\\
\begin{acknowledgments}
We would like to thank T.~Ohl for useful comments. The research of
F.P. is supported by Research Training Group 1147 ``{\it Theoretical
  Astrophysics and Particle Physics}''of Deutsche Forschungsgemeinschaft.
G.S. is supported by the Federal Ministry of Education and Research (BMBF) under contract number 05HT6WWA.
\end{acknowledgments}


\begin{thebibliography}{00}

\bibitem{SU5}
H.~Georgi and S.~L.~Glashow, Phys. Rev. Lett. {\bf 32}, 438
  (1974); H.~Georgi, in {\it Proceedings of Coral Gables 1975, Theories and
  Experiments in High Energy Physics}, New York, 1975.

\bibitem{Pati:1974yy}
  J.~C.~Pati and A.~Salam,
  Phys.\ Rev.\  D {\bf 10}, 275 (1974)
  [Erratum-ibid.\  D {\bf 11}, 703 (1975)].

\bibitem{typeIseesaw}
  P.~Minkowski, Phys.\ Lett.\  B {\bf 67}, 421 (1977); T.~Yanagida, in {\it
  Proceedings of the Workshop on the Unified Theory and Baryon Number in the
  Universe}, KEK, Tsukuba, 1979; M.~Gell-Mann, P.~Ramond and R.~Slansky, in
  {\it Proceedings of the Workshop on Supergravity}, Stony Brook, New York,
  1979; S.~L.~Glashow, in {\it Proceedings of the 1979 Cargese Summer
  Institute on Quarks and Leptons}, New York, 1980.

\bibitem{typeIIseesaw}
M.~Magg and C.~Wetterich, Phys.\ Lett.\ B {\bf 94}, 61 (1980);
  R.~N.~Mohapatra and G.~Senjanovi\'c, Phys.\ Rev.\ Lett.\ {\bf 44},
  912 (1980); Phys.\ Rev.\ D {\bf 23}, 165 (1981); J.~Schechter and J.~W.~F.~Valle,
  Phys.\ Rev.\ D {\bf 22}, 2227 (1980); G.~Lazarides, Q.~Shafi and
  C.~Wetterich, Nucl.\ Phys.\ B {\bf 181}, 287 (1981).

\bibitem{Harrison:1999cf}
  P.~F.~Harrison, D.~H.~Perkins and W.~G.~Scott,
  Phys.\ Lett.\  B {\bf 458}, 79 (1999);
  P.~F.~Harrison, D.~H.~Perkins and W.~G.~Scott,
  Phys.\ Lett.\  B {\bf 530}, 167 (2002).

\bibitem{PMNS}
B.~Pontecorvo, Sov.\ Phys.\ JETP {\bf 6}, 429 (1957); Z.~Maki,
  M.~Nakagawa and S.~Sakata, Prog.\ Theor.\ Phys.\ {\bf 28}, 870 (1962).

\bibitem{A4}
  E.~Ma and G.~Rajasekaran,
  Phys.\ Rev.\  D {\bf 64}, 113012 (2001);
  K.~S.~Babu, E.~Ma and J.~W.~F.~Valle,
  Phys.\ Lett.\  B {\bf 552}, 207 (2003); M.~Hirsch {\it et al.}, Phys.\ Rev.\  D {\bf 69}, 093006 (2004).

\bibitem{T'}
  P.~H.~Frampton and T.~W.~Kephart,
  Int.\ J.\ Mod.\ Phys.\  A {\bf 10}, 4689 (1995);
  A.~Aranda, C.~D.~Carone and R.~F.~Lebed,
  Phys.\ Rev.\  D {\bf 62}, 016009 (2000);
  P.~D.~Carr and P.~H.~Frampton,
  arXiv:hep-ph/0701034; A.~Aranda,
  Phys.\ Rev.\  D {\bf 76}, 111301 (2007).

\bibitem{delta27}
  I.~de Medeiros Varzielas, S.~F.~King and G.~G.~Ross,
  Phys.\ Lett.\  B {\bf 648}, 201 (2007);
  C.~Luhn, S.~Nasri and P.~Ramond,
  J.\ Math.\ Phys.\  {\bf 48}, 073501 (2007);
  Phys.\ Lett.\  B {\bf 652}, 27 (2007).


\bibitem{Ma:2007ia}
E.~Ma, arXiv:0705.0327 [hep-ph];
G.~Altarelli, arXiv:0705.0860 [hep-ph].

\bibitem{CKM}
N.~Cabibbo, Phys.\ Rev.\ Lett.\ {\bf 10}, 531 (1963); M.~Kobayashi and T.~Maskawa, Prog.\ Theor.\ Phys.\ {\bf 49}, 652 (1973).

\bibitem{discreteGUTs}
 M.-C.~Chen and K.~T.~Mahanthappa, 
Phys.\ Lett.\  B {\bf 652}, 34 (2007); W.~Grimus and H.~Kuhbock,
Phys.\ Rev.\  D {\bf 77}, 055008 (2008);
  F.~Bazzocchi {\it et al.}, arXiv:0802.1693 [hep-ph];
 G.~Altarelli, F.~Feruglio and C.~Hagedorn, J.~High Energy Phys.~{\bf 0803}, 052 (2008).

\bibitem{qlc}
  A.~Y.~Smirnov, arXiv:hep-ph/0402264;
  M.~Raidal,
  Phys.\ Rev.\ Lett.\  {\bf 93}, 161801 (2004);
  H.~Minakata and A.~Y.~Smirnov, Phys.\ Rev.\  D {\bf 70}, 073009 (2004).

\bibitem{Plentinger:2006nb}
  F.~Plentinger, G.~Seidl and W.~Winter,
  Nucl.\ Phys.\  B {\bf 791}, 60 (2008).

\bibitem{Plentinger:2007px}
  F.~Plentinger, G.~Seidl and W.~Winter,
  Phys.\ Rev.\  D {\bf 76}, 113003 (2007).

\bibitem{Plentinger:2008up}
  F.~Plentinger, G.~Seidl and W.~Winter, J.~High
  Energy Phys.~{\bf 0804}, 077 (2008).

\bibitem{Cacciapaglia:2006tg}
  G.~Cacciapaglia, C.~Csaki, C.~Grojean and J.~Terning,
  Phys.\ Rev.\  D {\bf 74}, 045019 (2006).

\bibitem{Agashe:2007jb}
  K.~Agashe, A.~Falkowski, I.~Low and G.~Servant,  J.~High
  Energy Phys.~{\bf 0804}, 027 (2008);
  C.~D.~Carone, J.~Erlich and M.~Sher,
  arXiv:0802.3702 [hep-ph].

\bibitem{Kawamura:2000ev}
  Y.~Kawamura,
  Prog.\ Theor.\ Phys.\  {\bf 105}, 999 (2001);
  G.~Altarelli and F.~Feruglio,
  Phys.\ Lett.\  B {\bf 511}, 257 (2001);
  A.~B.~Kobakhidze, Phys.\ Lett.\  B {\bf 514}, 131 (2001);
  A.~Hebecker and J.~March-Russell,
  Nucl.\ Phys.\  B {\bf 613}, 3 (2001);
  L.~J.~Hall and Y.~Nomura,
  Phys.\ Rev.\  D {\bf 66}, 075004 (2002).

\bibitem{Kaplan:2001ga}
  D.~E.~Kaplan and T.~M.~P.~Tait,
  J. High Energy Phys. {\bf 0111}, 051 (2001).

\bibitem{Froggatt:1978nt}
  C.~D.~Froggatt and H.~B.~Nielsen,
  Nucl.\ Phys.\  B {\bf 147}, 277 (1979).

\bibitem{Nomura:2001tn}
  Y.~Nomura,
  Phys.\ Rev.\  D {\bf 65}, 085036 (2002).

\bibitem{Georgi:1979df}
 H.~Georgi and C.~Jarlskog, Phys.\ Lett.\ B{\bf 86}, 297 (1979).

\bibitem{Arason:1991hu}
H.~Arason {et al.}, Phys.\ Rev.\ Lett.\ {\bf 67}, 2933 (1991);
    H.~Arason {et al.}, Phys.\ Rev.\ D {\bf 47}, 232 (1993).

\bibitem{Ayres:2004js}
  D.~S.~Ayres {\it et al.}  [NO$\nu$A Collaboration],
  arXiv:hep-ex/0503053; Y.~Hayato {\it et al.}, Letter of Intent.

\bibitem{qlcPatiSalam}
 S.~Antusch, S.~F.~King and
  R.~N.~Mohapatra, Phys. Lett. B {\bf 618}, 150 (2005).

\bibitem{Winter:2007yi}
  W.~Winter,
  Phys.\ Lett.\  B {\bf 659}, 275 (2008).

\bibitem{protondecay}
K.~S.~Babu and S.~M.~Barr, Phys.\ Rev.\ D {\bf 48}, 5354 (1993);
K.~Kurosawa, N.~Maru and T.~Yanagida, Phys.\ Lett.\  B {\bf 512}, 203 (2001).

\bibitem{Krauss:1988zc}
  L.~M.~Krauss and F.~Wilczek,
  Phys.\ Rev.\ Lett.\  {\bf 62}, 1221 (1989).

\bibitem{inpreparation}
F.~Plentinger and G.~Seidl, {\it in preparation}.

\end{thebibliography}
\end{document}